\documentclass[review]{elsarticle}
\usepackage{lineno,hyperref}
\usepackage[flushleft]{threeparttable}
\usepackage{amssymb}
\usepackage{float}
\usepackage[deletedmarkup=sout]{changes}
\usepackage{makecell}

\usepackage{nomencl} 
\makenomenclature

\journal{}

\begin{document}

\begin{frontmatter}

\title{A review of flywheel energy storage systems: state of the art and opportunities}

\author[mymainaddress,mysecondaryaddress]{Xiaojun Li\corref{mycorrespondingauthor}}

\cortext[mycorrespondingauthor]{Corresponding author}
\ead{tonylee2016@gmail.com}
\author[mymainaddress]{Alan Palazzolo}

\address[mymainaddress]{Dwight Look College of Engineering, Texas A\&M University, College Station, Texas, 77840, USA}
\address[mysecondaryaddress]{Gotion Inc, Fremont, CA, 94538, USA}

\begin{abstract}
Thanks to the unique advantages such as long life cycles, high power density, minimal environmental impact, and high power quality such as fast response and voltage stability, the flywheel/kinetic energy storage system (FESS) is gaining attention recently. There is noticeable progress in FESS, especially in utility, large-scale deployment for the electrical grid, and renewable energy applications. This paper gives a review of the recent developments in FESS technologies. Due to the highly interdisciplinary nature of FESSs, we survey different design approaches, choices of subsystems, and the effects on performance, cost, and applications. This review focuses on the state of the art of FESS technologies, especially those commissioned or prototyped. We also highlighted the opportunities and potential directions for the future development of FESS technologies.
\end{abstract}

\end{frontmatter}



\nomenclature{FESS}{Flywheel energy storage system}
\nomenclature{ESS}{Energy storage system} 

\printnomenclature[1in]

\section{Introduction}
In the past decade, considerable efforts have been made in renewable energy technologies such as wind and solar energies. Renewable energy sources are ideal for replacing fossil fuels to provide sustainable and clean energies. Besides, they are more available globally, where electrical shortages are frequent due to poor infrastructure. However, wind and solar power's intermittent nature prevents them from being independent and reliable energy sources for micro-grids. Energy storage systems (ESS) play an essential role in providing continuous and high-quality power. ESSs  store intermittent renewable energy to create reliable micro-grids that run continuously and efficiently distribute electricity by balancing the supply and the load \cite{Dunn2011}. The existing energy storage systems use various technologies, including hydroelectricity, batteries, supercapacitors, thermal storage, energy storage flywheels,\cite{Rimpel2021} and others. Pumped hydro has the largest deployment so far, but it is limited by geographical locations.  Primary candidates for large-deployment capable, scalable solutions can be narrowed down to three: Li-ion batteries, supercapacitors, and flywheels.

The lithium-ion battery has a high energy density, lower cost per energy capacity but much less power density, and high cost per power capacity. This explains its popularity in applications that require high energy capacities and are weight-sensitive, such as automotive and consumer electronics.  Comparing to batteries, both flywheel and supercapacitor have high power density and lower cost per power capacity. The drawback of supercapacitors is that it has a narrower discharge duration and significant self-discharges. Energy storage flywheels are usually supported by active magnetic bearing (AMB) systems to avoid friction loss. Therefore, it can store energy at high efficiency over a long duration. Although it was estimated in \cite{SCHMIDT201981} that after 2030, li-ion batteries would be more cost-competitive than any alternative for most applications. FESSs are still competitive for applications that need frequent charge/discharge at a large number of cycles. Flywheels also have the least environmental impact amongst the three technologies, since it contains no chemicals. It makes FESS a good candidate for electrical grid regulation to improve distribution efficiency and smoothing power output from renewable energy sources like wind/solar farms. Besides, because of their high power density and fast response time, typical applications of FESSs also include uninterrupted power service (UPS), hybrid locomotives, and power pulsation.

FESSs are introduced as a form of mechanical ESS in several books\cite{8004733,Rimpel2021}. Several review papers address different aspects of FESS researches \cite{Arani2017,MousaviG2017}. Many have focused on its application in renewable energies \cite{Arani2017}, especially in power smoothing for wind turbines\cite{Sebastian2012a}. There is also one investigation into the automotive area \cite{Hedlund2015}. These reviews have a strong emphasis on applications and grid integration or market overview/outlook\cite{WICKI20171118}.
Nevertheless, there is less review focusing on the technological aspects. Since FESS is a highly inter-disciplinary subject, this paper gives insights such as the choice of flywheel materials, bearing technologies, and the implications for the overall design and performance.  For the application survey, we focus on the FESS systems that have been commissioned or at least have completed a prototype system. \cite{app7030286,8004733} also give overviews of the main components and the related technologies for FESS. But they have less information regarding new trends and future directions.  This review focuses on the state-of-art of FESS development, such as the rising interest and success of steel flywheels in the industry. In the end, we discuss areas with a lack of research and potential directions to advance the technology.

\section{Working principles and technologies}

\begin{figure}[ht]
\centering
\includegraphics[width=10cm]{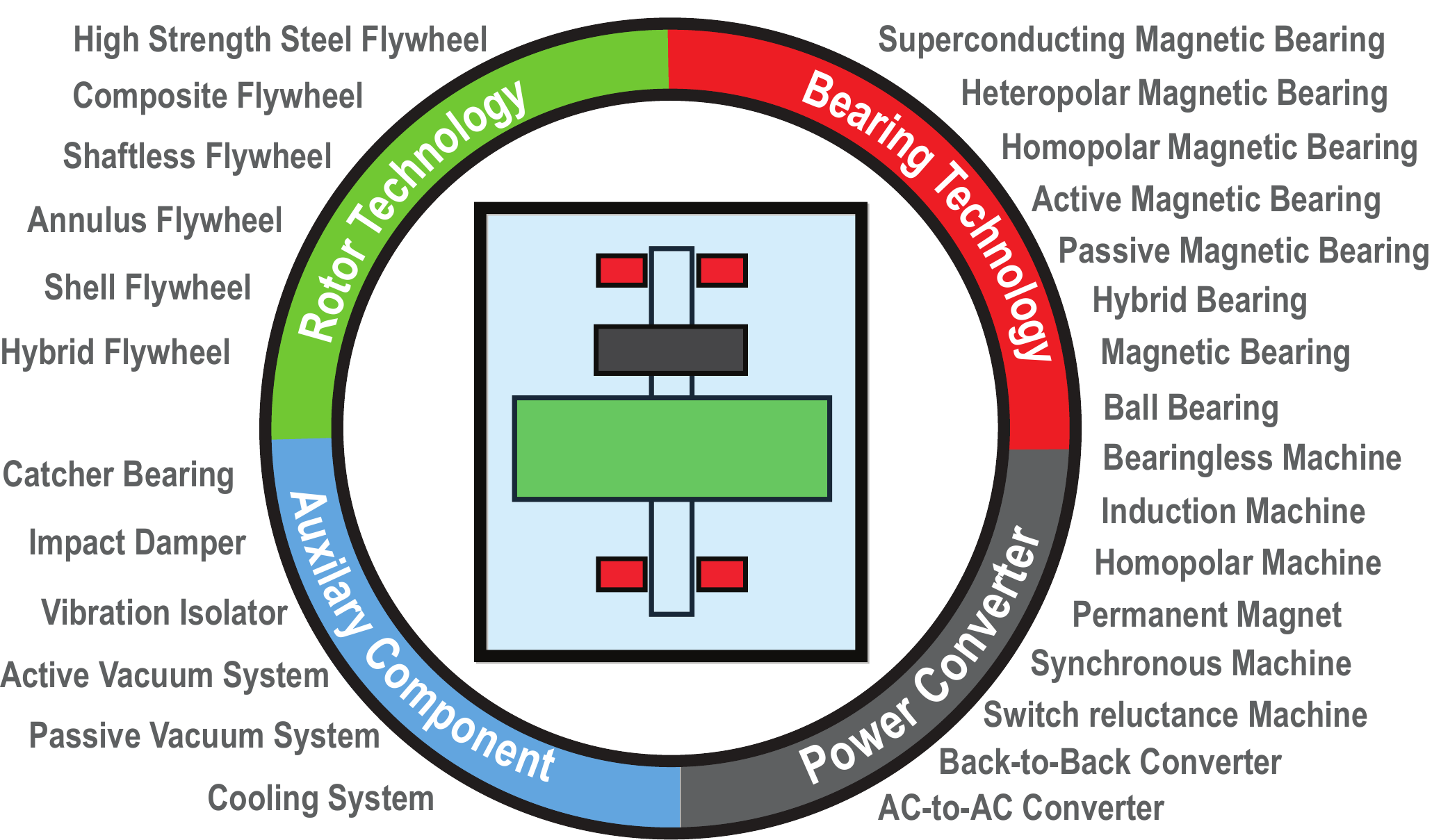}
\caption{An overview of system components for a flywheel energy storage system.}\label{sys_comp_circle}
\end{figure}

\subsection{Overview}

\begin{figure}[bt]
\centering
\includegraphics[width=8cm]{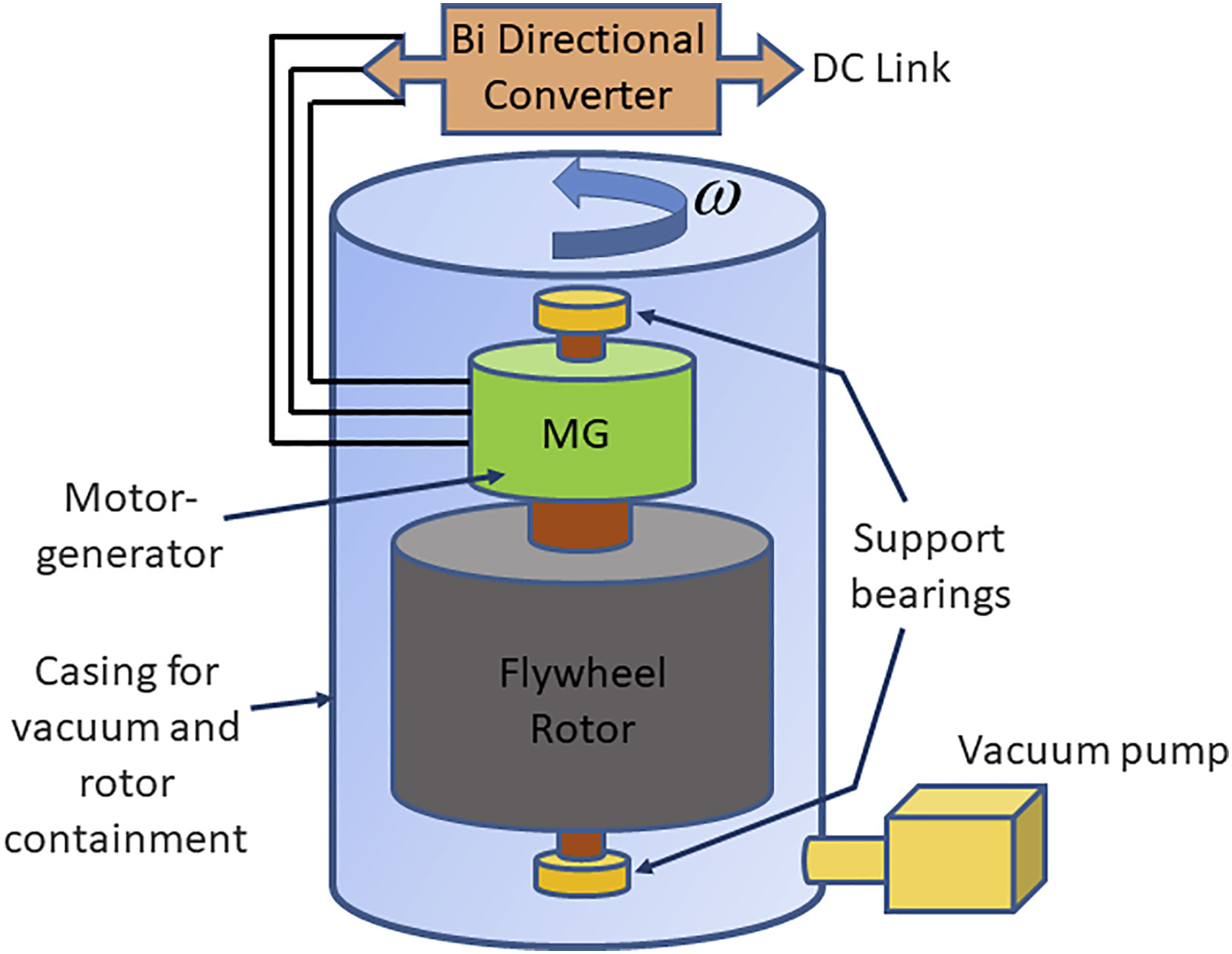}
\caption{A typical flywheel energy storage system \cite{Pullen2019TheStorage}, which includes a flywheel/rotor, an electric machine, bearings, and power electronics.}\label{typical}
\end{figure}

\begin{figure}[bt]
\centering
\includegraphics[width=8cm]{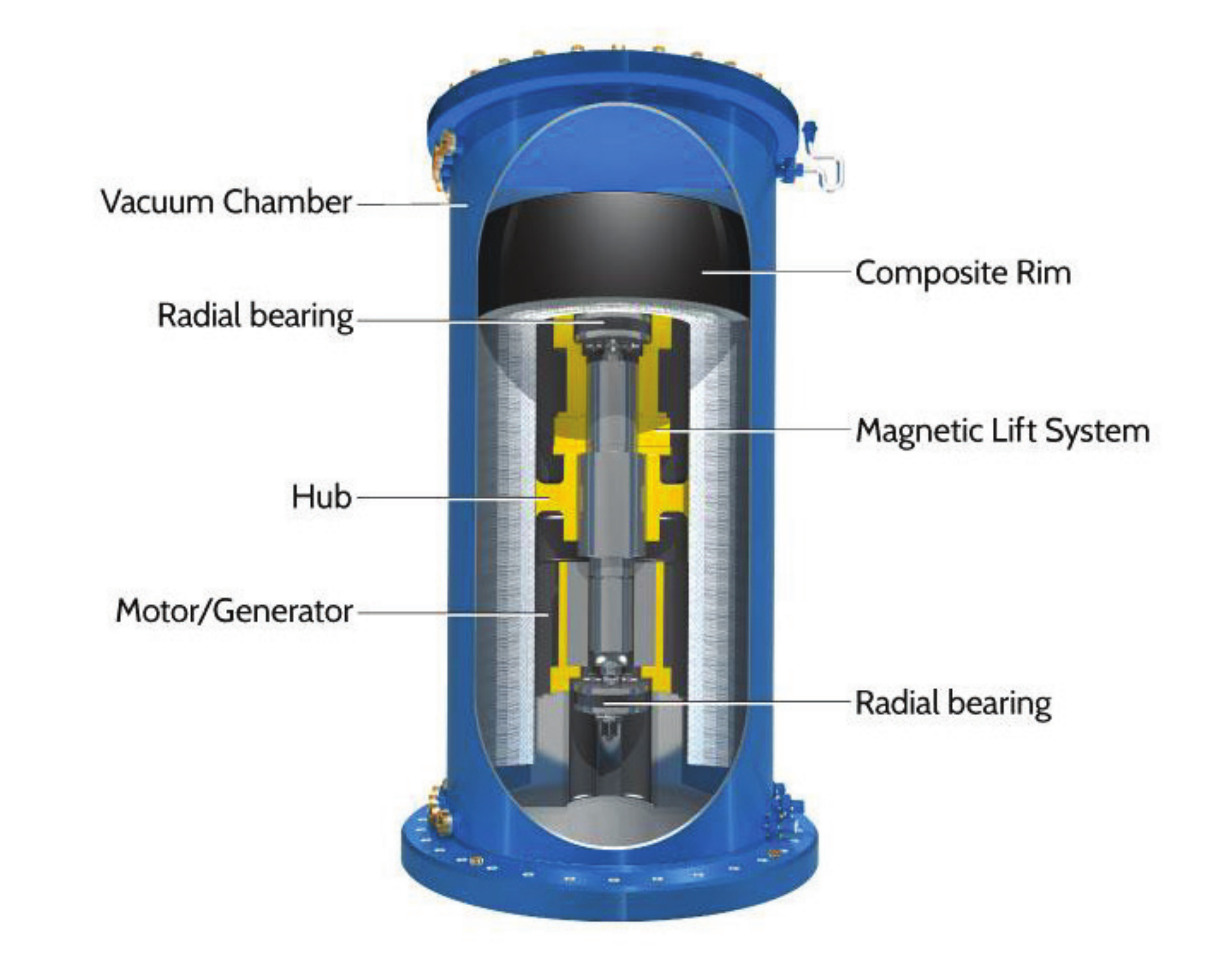}
\caption{The Beacon Power Flywheel \cite{Energy2013}, which includes a composite rotor and an electric machine, is designed for frequency regulation.}\label{beacon}
\end{figure}

Unlike the electrochemical-based battery systems, the FESS uses an electro-mechanical device that stores rotational kinetic energy (\(E\)) \nomenclature{$E$}{kinetic energy}, which is a function of the rotational speed (\(\omega\)) \nomenclature{$\omega$}{Flywheel's rotational speed} and the rotor's primary moment of inertia (\(I_p\)) \nomenclature{$I_p$}{Flywheel's primary moment of inertia}:

\begin{equation}\label{energy}
E = \frac{1}{2}I_p\omega^2.
\end{equation}

 Fig.~\ref{sys_comp_circle} has been produced to illustrate the flywheel energy storage system, including its sub-components and the related technologies. A FESS consists of several key components:1) A rotor/flywheel for storing the kinetic energy. 2) A bearing system to support the rotor/flywheel. 3) A power converter system for charge and discharge, including an electric machine and power electronics. 4) Other auxiliary components. As an example, the structure of a typical FESS is depicted in Fig.~\ref{typical}. 

\begin{figure}[t]
\centering
\includegraphics[width=10cm]{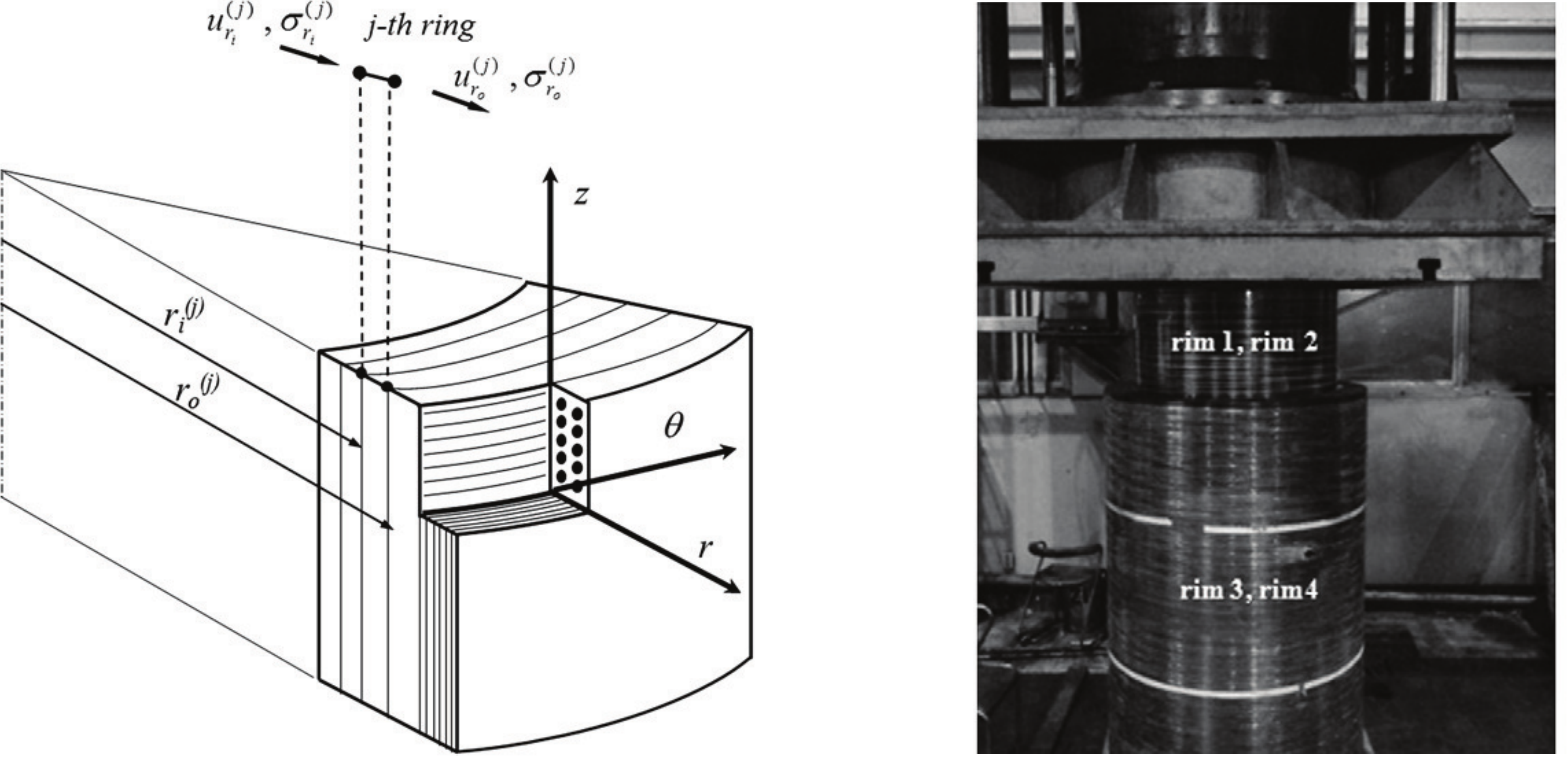}
\caption{Multi-rim composite flywheel and the shrink-fit process \cite{Ha2012}}\label{rotoroptimize}
\end{figure}

\begin{figure}[ht]
\centering
\includegraphics[width=8cm]{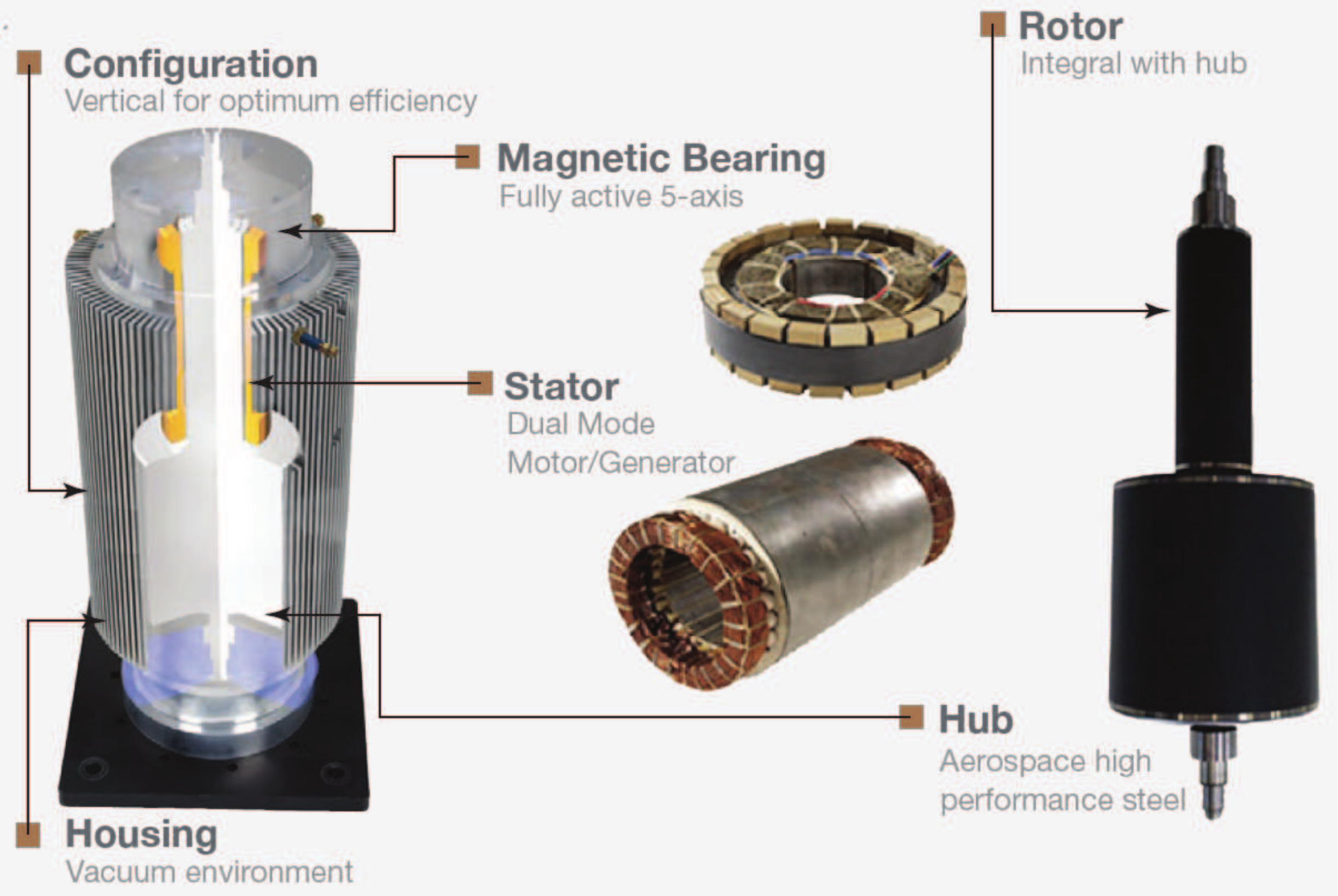}
\caption{ Calnetix/Vycon Flywheel \cite{Calnetix}, which includes a steel flywheel and an electric machine, is designed for UPS.}\label{vycon_fw}
\end{figure}

 To achieve a higher energy capacity, FESSs either include a rotor with a significant moment of inertia or operate at a fast spinning speed. Most of the flywheel rotors are made of either composite or metallic materials. For example, the FESS depicted in  Fig.~\ref{beacon} includes a composite flywheel rotor \cite{Energy2013}, whose operational speed is over 15,000 RPM. When spinning, the rotor is supported by operational bearings. The bearings can be either mechanical or magnetic. Magnetic bearings are preferred for minimal standby loss and maintenance requirements. For a composite flywheel, such as the one \cite{Ha2012} depicted in Fig.~\ref{rotoroptimize}, the rotor is usually carried by a metallic shaft that is magnetically permeable so that it can work with magnetic bearings and the motor/generator. Like the one depicted in Fig.~\ref{vycon_fw}, the shaft can be integrated with the rotor for a steel flywheel. A FESS also includes an energy converter. A mainstream choice is an electric machine like a motor/generator, such as the devices depicted in Fig.~\ref{vycon_fw}. The motor/generator converts the kinetic energy to electricity and vice versa. Alternatively, magnetic or mechanical gears can be used to directly couple the flywheel with the external load. To reduce standby loss, the flywheel rotor is often placed in a vacuum enclosure. Other auxiliary components include a vacuum pump, catcher bearings, and a cooling system.

\subsection{Flywheel/Rotor}

The flywheel (also named as rotor or rim) is the essential part of a FESS. This part stores most of the kinetic energy during the operation. As such, the rotor's design is critical for energy capacity and is usually the starting point of the entire FESS design. The following equations\cite{Genta1985} describe the energy capacity of a flywheel:
\begin{eqnarray}
\frac{E}{m} &= \alpha'\alpha''\alpha'''K\sigma/\rho \\
\frac{E}{v} &= \alpha'\alpha''\alpha'''K{\sigma}
\end{eqnarray}
where $\alpha'$ is the safety factor, $\alpha''$ the depth of discharge factor, $\alpha'''$ the ratio of rotating mass to the total system mass, $\sigma$ the material’s tensile strength\nomenclature{$\sigma$}{Flywheel's tensile strength}, $K$ the shape factor \nomenclature{$K$}{Shape factor}, and $\rho$ the density\nomenclature{$\rho$}{Flywheel's density}. A rotor with lower density and high tensile strength will have higher specific energy(energy per mass), while energy density(energy per volume) is not affected by the material's density. Typically, the rotor is carried by a shaft that is subsequently supported by bearings. The shaft also acts as the rotating part of the motor/generator. The orientation of the rotor-shaft assembly can be horizontal or vertical. Two kinds of materials are often chosen in building the rotor: composite and metal.

\subsubsection{Composite flywheel}
Research in composite flywheel design has been primarily focused on improving its specific energy. There is a direct link between the material's strength-to-mass density ratio and the flywheel's specific energy. Composite materials stand out for their low density and high tensile strength. Since they are anisotropic, composite materials have higher longitudinal tensile strength but much weaker radial tensile strength, the latter of which limits their energy capacity. Shrink-fitting multiple thin composite rims can improve this shortcoming by reducing stresses in the radial direction. The popular design criterion for composite flywheels is the Tsai-Wu failure criterion \cite{TSWU-COMPOSITE}. A composite flywheel usually includes several different materials such as carbon fiber, glass fiber, and epoxy. An optimization process is often carried out to find the optimal design considering rim thickness, shrink-fit allowances, and different material combinations \cite{Kim2014,app8081256,dUyar2020,Ramaprabha2021,RASTEGARZADEH2020}. Although composite materials can achieve a fairly high specific energy (50-
100 Wh/kg)\cite{Li2019}. It often needs a metallic shaft to interact with bearings and motor/generator, resulting in lower specific energy overall. When considering the whole flywheel, one of the composite prototypes \cite{Thelen2003} reached 11.7 Wh/kg.

\subsubsection{Steel flywheel}
Historically, steel flywheel was considered "low-speed" and "older" technology associated with high-loss mechanical bearing. There is less research in the steel/isotropic flywheel design \cite{REYNA2013,Kale2018}. These works focus on improving the specific energy and energy density by finding the optimal geometric profile or utilizing a novel configuration. Recently, steel flywheels are regaining many interests due to their advantages, such as low cost, easy fabrication, and better recyclability. Section \ref{steelfw} gives a more detailed discussion of the recent development of FESS based on high-strength steel flywheel.

\subsection{Operational Bearings}
Operational bearings are the set of bearings that support the rotor when it is under normal operation. One of the features of a modern FESS is the use of Magnetic Bearings (MB)\nomenclature{MB}{Magnetic Bearings}. MB allows the rotor being spinning without physically contacting any components to eliminate the friction loss, which is inevitable for mechanical bearings. For mechanical bearings, such as ball bearings, the power loss is roughly proportional to the rotor’s spinning speed. It prevents the FESS from running at a higher speed. Fluid-film bearings may have less power loss, but they need an extra lubrication system, making them inapplicable in a vacuumed FESS. A summary of different FESS bearing technologies is given in Table~\ref{FESS-bearing}. Notice that the homopolar AMB in the table refers to the commonly used, PM-biased homopolar AMB.

\subsubsection{Magnetic bearing}
The magnetic bearing of a FESS can be either active or passive. An active magnetic bearing (AMB)\nomenclature{AMB}{Active Magnetic Bearings} requires power electronics and a feedback controller. It can be homopolar, which means it has permanent magnets (or bias current) to provide the bias flux, or heteropolar, which does not include bias fluxes. Many commercial MBs are heteropolar AMBs due to the lower cost. Numerous research works in AMB have been published over the years. Schweitzer et al.\cite{Malsen} give a good overview of the magnetic bearing technology. For AMBs, control is an essential aspect. Control algorithms \cite{Li2018,Peng2018CompleteCompensation} are used to tackle issues like system couplings, gyroscopic effects, and synchronous vibrations. Passive magnetic bearings do not require a feedback control but have more complex designs than AMBs \cite{Filion2013Reduced-frictionTechnique}. Superconducting magnetic bearings are also extensively studied for flywheel energy storage \cite{Cansiz2019RotorApproximation,Morehouse2019,8559521,Miyazaki2016} for their superior performances. However, most of the designs are complicated and require cryogenic equipment.

\subsubsection{Hybrid bearing}\nomenclature{HB}{Hybrid bearing} 
Mechanical bearings, such as ball bearings, are rarely used to support a flywheel solely. However, they are used as a part of a hybrid bearing system\cite{Sanders2015}, together with magnetic bearings. Another typical hybrid bearing system combines passive and active magnetic bearing\cite{8508351,Basaran2019}. Han et al. \cite{Han2013} present a combination of passive magnetic bearings and active radial magnetic bearings. The passive magnetic bearings support the flywheel in the axial and provide stiffness in the tilting motion. In \cite{Basaran2019}, a radial repulsive magnetic bearing with less power consumption and design complexity, are proposed for a FESS.

\begin{table}[h]
\centering 
  \begin{threeparttable}
    \caption{FESS operational bearing technologies}\label{FESS-bearing}
    \begin{tabular}{lll}
    \hline
    Types  & Advantages & Disadvantages\\
    \hline
    Homopolar AMB       & \makecell[tl]{less loss \\ lower current } & \makecell[tl]{higher cost  \\ less robust \\ demagnetization} \\
    Heteropolar AMB     & \makecell[tl]{lower cost \\ more rubost} & \makecell[tl]{higher loss \\ higher current} \\
    HB                  & \makecell[tl]{less loss \\ lower cost} & \makecell[tl]{complex design \\ less mature} \\

    \hline  
    \end{tabular}
  \end{threeparttable}
\end{table}

\subsection{Energy converter}

\begin{figure}[th!]
\centering
\includegraphics[width=6cm]{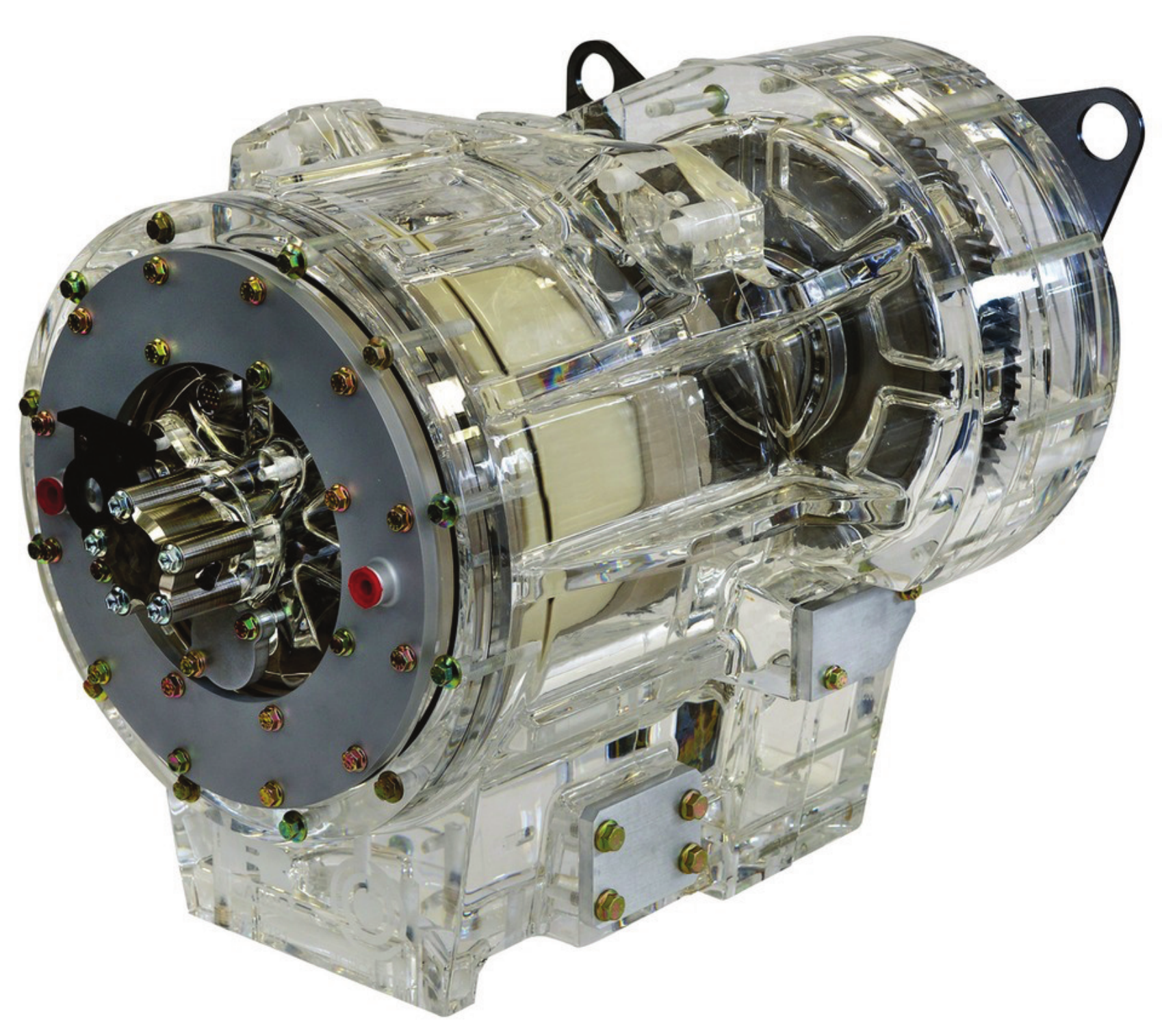}
\caption{Ricardo TorqStor \cite{SpinningWheel2015}, which includes a composite flywheel and magnetic gear, is designed for automotive applications.}\label{typicalflywheel}
\end{figure}

This section will discuss the energy converters for FESSs. For an overview of electromechanical energy conversion, the readers may refer to \cite{CANSIZ2018598}. In general, an electric machine is used to convert electrical energy into kinetic energy and vice versa. It is acting as a motor and generator. Permanent Magnet Synchronous Motors (PMSM)\nomenclature{PMSM}{Permanent Magnet Synchronous Motors} is one of the popular options for flywheel applications because of their high efficiency, high performance, and compact size. Other electric machines, such as induction motors (IM) or switch reluctance motors (SRM), are also used for flywheels. The M/G’s design, including the power density and current carrying capacity,  is crucial for the flywheel’s power rating. Apart from electric machines, the other option is to use magnetic gears (MGR) to link the flywheel with the external load. As depicted in Fig.~\ref{typicalflywheel}, magnetic gears do not require extra power electronics. They are relatively new for flywheels and will be covered in Section~\ref{compact}. A summary of different FESS electric machines is given in Table~\ref{FESS-Machine}. A more detailed comparison between PMSM, IM, and SRM can be found in \cite{Yang2015}.

\subsubsection{Permanent magnet synchronous machine}
A permanent magnet synchronous machine has high power density and efficiency. They are popular choices for FESS \cite{Anvari2018,Serpi2014,Lu2013}. Design considerations include magnet size, grade, number of poles. A significant design factor is that the machine needs to operate in a vacuum space, with radiation being the only mean of heat dissipation. One of the early works, Nagorny et al. \cite{Nagorny2005a} give an overview of the design considerations of PMSM for FESS applications. More recently, Schneider et al. \cite{Schneider2017} investigate the PMSM iron and copper loss based on an analytical model. The drawbacks of PMSMs are also related to the use of permanent magnets, which are subject to demagnetization. PMSM also suffers from idling losses. A PM machine is less rugged, less robust to temperature, and significantly more expensive than switched reluctance machines and induction machines. Kesgin et al. \cite{Korsunskiy_2020} discuss the progress and development trends in electric motor/generators employed in FESS, in which the potential of axial-flux permanent-magnet (AFPM) machines for FESS is highlighted.

\subsubsection{Induction machine}
Induction machines have lower costs and high robustness \cite{Yang2020}. However, standard induction machines are less efficient than PMSM. Arani et al. \cite{khodadoost} present the modeling and control of an induction machine-based flywheel energy storage system for frequency regulation after micro-grid islanding. Mir et al. \cite{Mir2018} present a nonlinear adaptive intelligent controller for a doubly-fed-induction machine-driven FESS. The control is for mitigating the intermittency in wind power injection and enhancing the transient stability of the connected power system thereby. In \cite{app9214537}, Soomro et al. give a model to access the operational and standby losses of a squirrel-cage induction machine-based FESS.

\subsubsection{Switch reluctance machine}
A Switch reluctance machine\cite{9296247,8613857,9160512} (SRM)\nomenclature{SRM}{Switch reluctance machine} requires no PM but offers similar or better efficiency under higher speeds when comparing to induction machines. According to \cite{Liaw2018}, SRM also has high acceleration capability, no cogging torque, high efficiency, a simple converter circuit, and having the fault-tolerant ability. But SRM is considered less mature and unproven than PMSM and induction machines.

\subsubsection{Bearingless machine}\nomenclature{BM}{Bearingless machine}\label{bearingless}
FESS may use a bearingless machine that combines two functionalities (MG and MB) into one device\cite{Yuan2016,Sun2018,9296247,Yang2021}. For bearingless machine design, both the suspension force and torque performance needs to be considered \cite{Sun2018}. In \cite{8613857}, a five-phase bearingless flux-switching permanent magnet machine is presented. The machine's parameters are optimized to improve both torque and suspension force with increased amplitude and minor fluctuation.

Other design options include brushless direct current machine (BLDC)\nomenclature{BLDC}{Brushless direct current machine} \cite{8369377,9012455} and AC homopolar machine\cite{7134770,en12010086}. BLDC has high power density, high efficiency, and compact form factor\cite{MousaviG2017}. The homopolar machine has a simple but robust structure and low idling loss.

\renewcommand\theadalign{bc}
\renewcommand\theadfont{\bfseries}
\renewcommand\theadgape{\Gape[4pt]}
\renewcommand\cellgape{\Gape[4pt]}

\begin{table}[h!]
\centering 
  \begin{threeparttable}
\caption{FESS electric machine designs}\label{FESS-Machine}
\begin{tabular}{lll}
\hline
Types  & Advantages & Disadvantages\\
\hline
PMSM  & \makecell[tl]{higher power density \\ high efficiency\\ small form factor} & \makecell[tl]{higher cost \\ demagnetization \\ less robust \\higher idling loss} \\
IM    & \makecell[tl]{less cost \\ more rugged \\ simple construction} & \makecell[tl]{lower power density  \\ less efficiency \\ higher slip ratio} \\
SRM   & \makecell[tl]{no demagnetization \\ more rugged \\simple construction} & \makecell[tl]{ complex control \\ less mature} \\
BM    & \makecell[tl]{high integration level} & \makecell[tl]{higher cost and complexity \\ complex control} \\
MGR    & \makecell[tl]{no power electroniccs\\ more rugged \\simple construction} & \makecell[tl]{ less power density\\no active control \\less mature} \\

\hline  
\end{tabular}
  \end{threeparttable}
\end{table}

\subsubsection{Power electronics}
Power electronics can be viewed as an interface between the electric machine and the electrical load/supply. Different designs and control methods are proposed to achieve high power/current capability with fewer disturbances for the grid. A typical design is using a back-to-back converter that includes two voltage source controllers (VSC)\nomenclature{VSC}{Voltage source controllers}. The VSCs switch their roles between rectifiers and inverters to realize the transformation between charge and discharge modes. The current carrying capacity of the VSC is also a critical factor in determining the FESS’s power rating.  Bernardinis et al.\cite{Bernardinis2017} design a high-efficiency inverter. The inverter is tested at 20kHz and achieved 98.8\% efficiency at 60kW. In \cite{Gurumurthy2016}, a new topology for a bidirectional converter is presented. This converter targets zero switching power loss for the buck and boost modes. The scheme is verified by a 4kW, 340V prototype flywheel, where a 2.5-3.5\% power saving is observed.

Controller design for power converters is also a major topic. Zhang et al.\cite{Zhang2018} propose a method using direct voltage control for the DC-link without the intermediate current loop. Testing results show reduced tracking error and steady-error at different spinning speeds. In \cite{GAMBOA2019214}, an Input-Output Linearization (IOL) AC voltage controller is presented for a FESS supplied by an AC-AC matrix converter. The system can compensate for the critical load voltage without noticeable delays, and voltage undershoots/overshoots. It also overcomes the input/output coupling problem of matrix converters. Multilevel converters are proposed for FESS as well. Murayama et al. \cite{8293859} present a modular multilevel cascade converter (MMCC) for FESS. The converter is intended to achieve rapid current control without creating a significant disturbance on the grid. Applications are plasma control, particle accelerators, or medical use.

\subsection{Auxiliary components}

\subsubsection{Catcher bearing}
Catcher Bearings (CB)\nomenclature{CB}{Catcher bearings} are sometimes referred to as touchdown or backup bearings. They are auxiliary bearings used as a backup in case of MB failures caused by power shortage or excessive external disturbances. CBs are not designed to provide operational support in place of MBs. Ball bearings are typically used as CBs. In \cite{JIN2015253}, Jin et al. analyze the thermal structure of two types of catcher bearing. Both theoretical and experimental results show that the double-decker catcher bearing (DDCB) is more resistant to temperature rise than the single-decker catcher bearing (SDCB). 

\subsubsection{Other components}
Many of the housing/casing designs of FESSs also include a vacuum enclosure to reduce the windage loss when the flywheel is rotating at high speed. For mobile applications, the housing structure needs to be optimized to reduce its overall weight. It also needs to provide vibration adsorptions to prevent the FESS from failures caused by excessive external vibrations. The flywheel that operates in a vacuum enclosure may also include other components such as an air pump for maintaining its vacuum status and an active cooling system for the MB and M/G.

\section{Applications}


The applications of FESSs can be categorized according to their power capacity and discharge time. Recently developed FESSs have lower costs and lower losses. They can work for multiple hours \cite{M32} instead of just several minutes or seconds. Besides, FESSs boast advantages like long life cycles, fast responses, and less sensitivity towards temperature and humidity. This gives FESSs the potential to replace electrochemical batteries in the grid and renewable energy applications.  This section will focus on the systems that have been commissioned or prototyped. We have summarized a list of FESSs research groups in Table~\ref{fw-research}. The list of commercial flywheel systems is given in Table~\ref{fw-comm}.

\subsection{Utility application}

\subsubsection{Frequency regulation}
Frequency regulation is one of the driving forces for FESS research and development. Most utility electricity is generated by gas turbines operating at a specific speed range for high efficiency. However, the load of a power grid is not constant. Minute-to-minute variability is caused by the random turning on and off of millions of individual loads. It is challenging to balance the generation and load in real-time \cite{Kirby2004}. Whenever the load exceeds the generation, more kinetic energy is drawn from the turbine, causing it to slow down. Subsequently, the grid frequency deviates from its nominal value. Only a few tenths of a hertz of frequency deviation can cause damage to valuable equipment. Energy storage systems act as virtual power plants by quickly adding/subtracting power so that the line frequency stays constant. FESS is a promising technology in frequency regulation for many reasons. Such as it reacts almost instantly, it has a very high power to mass ratio, and it has a very long life cycle compared to Li-ion batteries. The main advantage is the long life cycles, which significantly lowers the long-term operational cost. Beacon Power \cite{Energy2013} is one of the early companies that focuses on FESS technology for grid applications. They have successfully commissioned a 20 MW FESS plant in Pennsylvania. The rotor is made of carbon fiber, which operates at 16,000 RPM. It also has a 175,000 life cycle. Helix Power \cite{HelixInc} is developing 1-MW and 90s FESS for grid application. The flywheel's steady-state power loss is less than 1\% of the rated power. Many research works focus on control. Mahdavi et al. \cite{9154722} presents an enhanced frequency control system and its experimental verification for a FESS to reduce the frequency variations of the microgrid. In \cite{Yao2017}, a fuzzy, PD-based frequency regulation control strategy for wind-power and FESS system proposed to enhance the frequency regulation capability of direct-drive permanent magnet synchronous generator (PMSG)-based wind-power generation system. 


\subsubsection{Renewable energy integration}
Renewable energy is another area with high research activities. Since wind is unpredictable, a wind turbine has fluctuating power output. For solar PVs (Photovoltaics)\nomenclature{PV}{Photovoltaics}, sun radiation varies within the daytime and is not available during the night. Many FESSs have been proposed to smooth the output and increase a wind turbine or solar farm's efficiency. Here, we do not intend to give yet another comprehensive survey in this field, which already has abundant reviews \cite{Sebastian2012a,Abdeltawab2016,MousaviG2017}. More recently, Gayathri et al. \cite{Gayathri2017} present a nonlinear controller for smoothing the power output from a wind turbine. Azizimoghaddam et al. \cite{9258391} proposes a model that includes an integrated model including both power network and FESS parameters. The model is used for optimization to achieve optimum dynamic performance. Hitachi ABB has installed a 2 MW flywheel system for 15,000 inhabitants on Kodiak Island, which plans to run entirely on renewable energy. It was reported \cite{Fairly2015} that the flywheel system is "the first line of defense against varying power flows from wind turbines, relieving a 3-MW battery system that is wearing out faster than expected". In \cite{RAMLI2015398}, a flywheel is used to store excess energy from a PV-diesel hybrid energy system. Its economic and environmental benefits are studied.

\subsubsection{Uninterruptible power system}
Many of the commercial flywheel systems are developed and marketed for UPS\nomenclature{UPS}{Uninterruptible power supply} applications. The key advantages of flywheel-based UPS include high power quality, longer life cycles, and low maintenance requirements. Active power Inc. \cite{ActiveInc} has developed a series of flywheels capable of 2.8 kWh and 675 kW for UPS applications. The flywheel weighs 4976 kg and operates at 7700 RPM. Calnetix/Vycons’s VDC \cite{CalnetixVDC} is another example of FESS designed for UPS applications. The VDC's max power and max energies are 450 kW and 1.7 kWh. The operational range is between 14,000 RPM and 36,750 RPM. Lashway et al. \cite{Lashway2016} have proposed a flywheel-battery hybrid energy storage system to mitigate the DC voltage ripple. Interestingly, flywheels are also used to provide backup power for nuclear power plants \cite{Abdussami2019Flywheel-basedPlant}.

\subsection{Transportation application}
\subsubsection{Automotive}
The FESSs are used in both vehicular and transportation applications. M. Hedlund et al. \cite{Hedlund2015} gives a review of FESS applications in automobiles. Its high power to mass ratio enables the FESS to replace conventional powertrain systems \cite{Li2019}. In \cite{Abrahamsson2014b}, an energy buffer storing up to 867 Wh is presented. It is primarily for utility vehicles in urban traffic. R. A. Smith and K. R. Pullen \cite{Read2015} present the optimization of a flywheel designed for braking energy recovery and acceleration for hybrid vehicles. The result is optimal flywheel size and depth-of-discharge for a particular vehicle to achieve a balance between high transmission efficiency and low system mass. In \cite{Korsunskiy_2020}, a simulation model is proposed to evaluate the dynamic qualities and efficiency of a heavy-duty transport vehicle equipped with a mechanical transmission, using a combined power plant with a FESS. Ershad et al. \cite{Ershad2021} propose a flywheel-based four-wheel-drive, a full-electric powertrain that significantly increases the overall performance and battery lifespan.


\subsubsection{Locomotives/trains}
Notable early work includes The University of Texas 2MW flywheel system as a part of the advanced locomotive propulsion system\cite{Thelen2003}. More recent developments include the REGEN systems \cite{Calnetix}. The REGEN model has been successfully applied at the Los Angeles (LA) metro subway as a Wayside Energy Storage System (WESS)\nomenclature{WESS}{Wayside Energy Storage System}. It was reported that the system had saved 10 to 18\% of the daily traction energy. The LA metro Wayside Energy Storage Substation (WESS) includes 4 flywheel units and has an energy capacity of 8.33kWh. The power rating is 2 MW. The analysis \cite{Solis} shows that "the WESS will save at least \$99,000 per year at the Westlake/MacArthur Park TPSS". The FESS is made of steel. The flywheel is also designed to be fully levitated by magnetic bearings. Its operational speed range is from 10,000 to 20,000 RPM. Flywheel is often applied in heavy-haul locomotive \cite{SPIRYAGIN2015607,Spiryagin2018}. For example, Spiryagin et al. \cite{SPIRYAGIN2015607} propose a simplified control strategy for a FESS-assisted heavy haul locomotive. The study concludes that "FESS can be a very good solution " because battery's limits on "specific power, cost efficiency and service lifetime".

\subsubsection{Marine}
FESSs have been designed as auxiliary parts of electrified ships to improve their power qualities \cite{Huynh2006,8433279,Langston2017,Hou2018}. As one of the early works, Huynh et al.\cite{Huynh2006} proposed a FESS design with low-loss magnetic bearings and a high-efficiency motor/generator. The FESS can output 500kW for 30s in high-duty mode and up to 2MW in pulse mode. More recently, Kumar et al. \cite{8433279} present the usage of a 50 kWh flywheel for a diesel-mechanical propelled tugboat. Test results show that with the adoption of variable speed operation of diesel generators, the flywheel offers 25.6\% fuel reduction. In \cite{Hou2018}, Hou et al. present a Battery-flywheel hybrid ESS to isolate load fluctuations from the shipboard network. Pulsed power applications on ships will be discussed in the following section. 

\subsection{Defense and aerospace}

\subsubsection{Pulsed power}
Owing to their high power density, FESSs have been used in Electromagnetic Launching systems (EMALS)\nomenclature{EMALS}{Electromagnetic Launching systems} and laser systems. As one of the early works, Swett et al.\cite{Swett2005} proposed a FESS for EMALS application. The system is designed to have a peak power output of 84.3 MW and an energy capacity of 126 MJ, equivalent to 35 kWh. In \cite{Gattozzi2015}, a simulation model has been developed to evaluate the performance of the battery, flywheel, and capacitor energy storage in support of laser weapons. FESSs also have been used in support of nuclear fusions. Rendell et al. \cite{Rendell2015} give a review of two Flywheel Generator Converters (FGCs) used by Joint European Torus (JET),  each flywheel supply 2600MJ (722 kWh) to their respective magnet load coils to supplement the 575MW (pulsed) grid supply. These flywheels have been in service for 30 years since 1983 and provided for approximately 85,000 JET pulses.

\subsubsection{Aerospace}
Many of the FESS research work in aerospace focus on replacing lion batteries with astronautical FESSs.  Satellites or space stations benefit from the flywheel’s high-power rating and long life cycle. The International Space Station has investigated the use of FESS by carrying out flight tests \cite{Edwards}.  FESS applications in satellite attitude control are a major topic in this field and will be covered in \ref{att-control}.
Although the high power density also gives FESS potentials in aeronautical applications, the authors have not noticed any research activity where FESSs are directly applied to an aeronautical aircraft. Consider the low specific energy, flywheel are not suitable to be used onboard. But they can be used as an ESS for aircraft take-off and landing.

\subsection{Research and industrial groups}
Various flywheel energy storage research groups\cite{Tang2012,Ha2012,Amiryar2020AnalysisSystems,Sheffield-2020,Miyazaki2016,Rupp2016,Hartl2015DesignRotor,Li2017h,Tsao2003,Abrahamsson2014b,Thelen2003,Kailasan2014} and industrial products \cite{ABBMicrogrid,ActiveInc,Sanders2015,Energy2013,McIver2010,CATUPS,Calnetix,Gerotor,HelixInc,piller,Powerthru,stornetic,Rotonix,GKN,velkess,Energiestro} are summarized in Table~\ref{fw-research} and Table~\ref{fw-comm}, which include the rotor materials, energy \& power density, storage duration, and applications.

\nomenclature{$P$}{Power rating}
\nomenclature{$\Delta{t}$}{Storage duration}

\begin{table}[h!]
  \begin{threeparttable}
\caption{A summary of FESS research groups}\label{fw-research}
\begin{tabular}{lrrrrrp{17mm}}
\hline
\textbf{Groups} & rotor &  $E$ & $P$ & $\Delta{t}$ & application\\
\hline
Beihang \cite{Tang2012}        & cm & - & - & - & Space \\
Hanyang U \cite{Ha2012}         & cm & 35 kWh & - & -  & - \\ 
London U \cite{Amiryar2020AnalysisSystems}   & st & 5 kWh & 10 kW & 30 min & - \\
Sheffield U \cite{Sheffield-2020} & - & 10 kWh & 500 kW& -  & Grid \\
Railway Tech. Res. Ins,t. \cite{Miyazaki2016}     & cm & 100 kWh & 300 kW& 20 min  & Renew \\
U of Alberta \cite{Rupp2016}            & - & - & - & - & Train \\
Technische U Wien \cite{Hartl2015DesignRotor}            & cm & 5k Wh & - & - & - \\ 
Texas A\&M U \cite{Li2017h}            & st & 100 kWh & 100 kW & 1hr & Grid \\
UC-Berkeley \cite{Tsao2003}                             & st  & 140 Wh & 30 kW & 16.8 s  & - \\
Uppsala U \cite{Abrahamsson2014b}        & cm & 867 Wh & - & - & - \\
UT-CEM \cite{Thelen2003}                                       & cm & 130 kWh & 2 MW & 3 min  & Train \\
Virginia U \cite{Kailasan2014}                                       & cm & 1 kWh & - & -  & - \\

\hline  
\end{tabular}
  \end{threeparttable}
    \begin{tablenotes}
      \small
      \item $\Delta{t}$ is based on reference or caculation when the rated power is given
      \item Rotor materials: cm - Composite; st - Steel; cn - Concrete; eg - E-glass
    \end{tablenotes}
\end{table}

\begin{table}
  \begin{threeparttable}
\caption{A summary of commercial FESS systems}\label{fw-comm}
\begin{tabular}{lrrrrrp{17mm}}
\hline
\textbf{manufacturer} & rotor &  $E$ & $P$ & $\Delta{t}$ & stage & application\\
\hline
Hitachi ABB \cite{ABBMicrogrid}        & -  & - & 2 MW & - & \checkmark & Wind \\
Active Power \cite{ActiveInc}          & st & 2.83 kWh & 675 kW & 15 s&  \checkmark & Various\\
Amber Kinetics \cite{Sanders2015}     & st & 32 kWh & 8 kW & 4 hrs & $\Uparrow$ & Various \\ 
Adaptive Balancing Power \cite{ADPW}   & cm & 12 kWh & 400 kW & 1.8 min & $\Rightarrow$ & Auto \\
Beacon Power \cite{Energy2013}       & cm  & 25 kWh & 100 kW & 15 min & \checkmark & Grid\\ 
Boeing \cite{McIver2010}             & cm  & 5 kWh & 3 kW & 1.7 hr & $\Uparrow$ & - \\ 
Caterpillar \cite{CATUPS}                 & -  & 5 kWh & 675 kW & 15s & \checkmark & UPS\\ 
Calnetix/Vycon \cite{Calnetix}     & st & 0.52 kWh & 125 kW & 15s & \checkmark & UPS \\ 
Dynamic Boosting System \cite{DBS}   & st & - & - & - & - & - \\
Gerotor \cite{Gerotor}               & - & 0.065 kWh & 50 kW & - & \checkmark & -\\
Kinetic Traction \cite{KNT}          & cm & - & 333 kW & - & $\Uparrow$ & Various\\
GKN Hybrid Power \cite{GKN}   & cm & 0.44 kWh & 120 kW & 13.2 s & $\Uparrow$ & Auto \\
Helix Power \cite{HelixInc}     & cm & 25 kWh & 1 MW & 90 s & $\Rightarrow$  & Grid\\
Levistor Flywheel \cite{LVF}    & st & 5 kWh & 100 kW& 3 min & $\Rightarrow$ & -\\
Oxto Energy \cite{OXE}    & - & 7.5 kWh & 65 kW & 7 min & $\Uparrow$ & Various \\
Piller Group \cite{piller}     & st  & 2.9 kWh& 625 kW & 15 s &  \checkmark & Various \\ 
Punch Flybrid \cite{PFL}   & st & 0.167 kWh & - & - & $\Uparrow$ & Various \\
Powerthru \cite{Powerthru}  & cm & .53 kWh & 101 kW & 10 - 25s & \checkmark & Defense \\ 
Ricardo TorqStor \cite{RCDO}          & cm & .056 kWh& 101 kW & - & $\Uparrow$  & Auto\\
Rotonix \cite{Rotonix}          & cm & 12 kWh& 1.1 MW & - & \checkmark & Various\\
Stornetic \cite{stornetic}      & cm & 3.6 kWh & 80 kW & 260s & \checkmark & Grid \\
Velkess* \cite{velkess}                        & eg & 15 kWh & 3 kW & - & $\Downarrow$ & Grid\\
Energiestro* \cite{Energiestro}               & cn & - & - & - & $\Rightarrow$ & Grid\\
\hline  
\end{tabular}
  \end{threeparttable}
    \begin{tablenotes}
      \small
      \item $\Delta{t}$ is based on reference or caculation when the rated power is given
      \item * unproven design, readers are suggested to view them with discretion.
      \item Rotor materials: cm - Composite; st - Steel; cn - Concrete; eg - E-glass
      \item Development stage : \checkmark commissioned; $\Uparrow$ prototyped; $\Rightarrow$ conceptual ; $\Downarrow$ discontinued
    \end{tablenotes}
\end{table}

\newpage

\section{Trends and future topics}

The current FESSs are not yet widely adopted as a popular energy storage solution. They have higher capital costs than electrochemical batteries \cite{Remillard2016,SCHMIDT201981}. For instance, Beacon Power’s flywheel costs almost ten times higher than a Li-ion battery system with similar energy capacity even though it can provide competitive cost per (kWh*cycles) considering the higher charge/discharge cycles. Compared to other technologies like batteries or supercapacitors, FESSs have “moving” parts, thus are considered to have higher uncertainty in failure modes. Composite flywheels are particularly susceptible to this shortcoming because of higher operational speeds and less predictable mechanical properties. 

Almost all the existing flywheel systems are designed for specific applications such as frequency regulation or UPS. They require specialized knowledge and techniques for manufacture, assembly, and maintenance, which prevents them from being produced in large quantities to reduce cost per unit. To address these issues, new efforts are made in different aspects of the technology. In the following, we discuss the emerging fields and potential opportunities for FESS technology.

\subsection{New Technologies}

\subsubsection{The comeback of high strength steel flywheels}\label{steelfw}

\begin{table}[b!] 
  \begin{threeparttable}
\caption{Comparison of different flywheel materials\cite{Bolund2007,XJL2018-TAMU}\label{steelfwtb}}
%
%
\begin{tabular}{lrrrrr}
\hline
Materials & Density &  Tensile strength & Specific Energy& Material cost\\
 & ($kg/m^3$) &  $MPa$ & $Wh/Kg$ & $\$/Kg$\\
\hline
Steel 4340 & 7700 & 1520 & 50 & 1 \\
E-glass 2000 & 100 & 1520 & 14 & 11 \\
S2-glass & 1920 & 1470 & 210 & 24.6 \\
Carbon T1000 & 1520 & 1950 & 350 & 101.8 \\
Carbon AS4C & 1510 & 1650 & 300 & 31.3 \\
\hline  
\end{tabular}
  \end{threeparttable}
\end{table}

Steel flywheels are often categorized as old and less efficient. It is no longer the case. Many recently developed FESSs, both by academia and the industry, are based on high-strength steel for competitive cost and broader availability. Composite materials are often chosen to make FESS flywheels for their low density and high tensile strength. Light-weight composite materials have a very high specific energy, which is crucial in aerospace or mobile applications research works \cite{Ha2012,HIROSHIMA2015304,Li2019} have claimed high specific energies between 50 to 150 Wh/kg. However, only the composite rim was included in the calculation. The metallic shaft, which is an essential component and has considerable mass, is normally neglected. Other components are also not considered. One of the composite-based FESS being successfully developed \cite{Thelen2003}, For example, it has a specific energy of 42KJ/kg, equivalent to only 11.7Wh/kg. The specific energy drops to 5.6Wh/kg when the whole system weight is included. 
\par
The cost of composite materials is significantly higher than steel too. The comparison of density, tensile strength, and costs between composite and steel is summarized in Table ~\ref{steelfwtb}. While T1000 has a lower density ( 20\% of steel’s) and higher tensile strength (26\% higher than steel), its cost is almost 100 times more. From the standpoint of cost, FESSs based on high-strength steels are more suitable for massive productions. More recently, there are several new flywheel prototypes made in high-strength steel \cite{Li2017h,Sanders2015,ActiveInc,CalnetixVDC}.
A comparative study \cite{Kale2018} also concluded that "for applications where the energy-per-cost is to be maximized, metals are superior to composite rotor materials." ground-based ESS applications are more sensitive to system space than weight. High-strength steel flywheels have a high energy density (volume-based energy) due to their high mass density.  Furthermore, they are superior to composite ones regarding thermal conductivity and design data availability, such as SN curves and fracture toughness. Therefore, High-strength steel flywheels are very suitable for fixed, ground-based, and large-capacity applications.

\subsubsection{New flywheel designs}

\begin{figure}[bt]
\centering
\includegraphics[width=10cm]{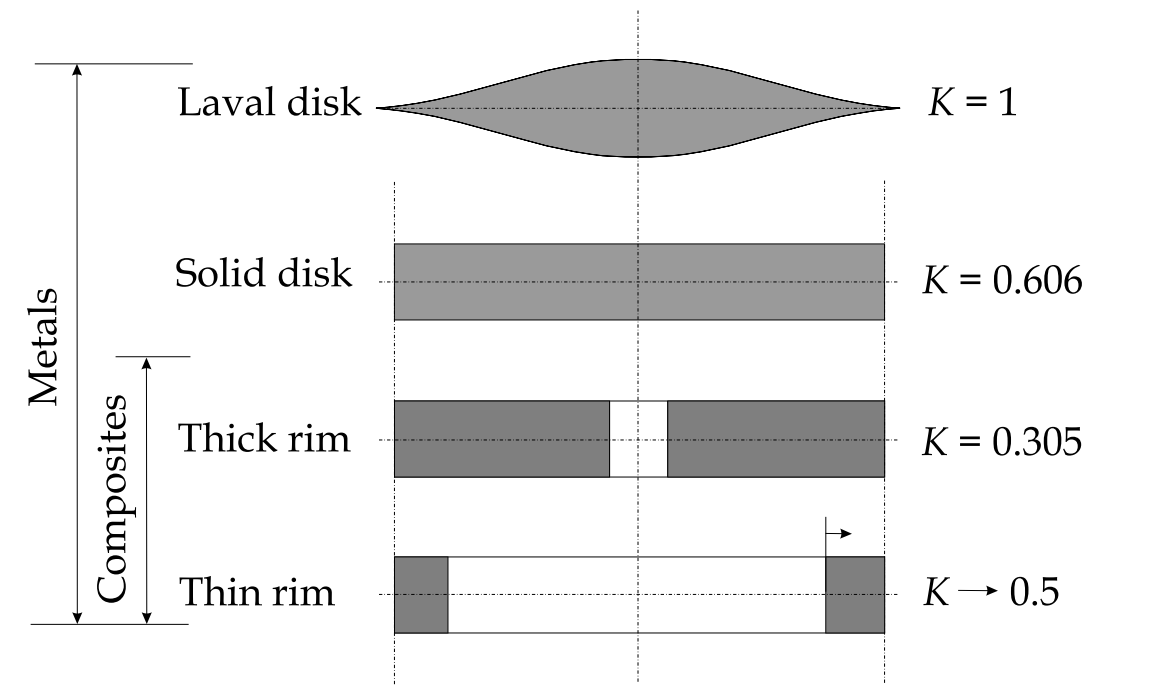}
\caption{Shape factors of typical flywheel designs\cite{Genta1985,Holm2003}\label{shape_factor}}
\end{figure}

Apart from pursuing higher spinning speed, raising the shape factor $K$ can also achieve higher specific energy and energy density. The shape factors of different flywheel designs are depicted in Fig.~\ref{shape_factor}. Conventional flywheels include an annulus rotor (thick rim in Fig.~\ref{shape_factor}). A shaft is shrink-fitted into its borehole, which increases the stress concentration. As a result, a conventional flywheel only has a shape factor of up to 0.3. A Laval disc \cite{Bolund2007} boasts an ideal shape factor of one. However, the geometry profile makes it very difficult to manufacture or suspend magnetically. Instead, a shaftless flywheel, which can be made in a single piece, has a shape factor close to 0.6, giving it almost a doubled specific density than the conventional design \cite{Li2017h,Li2015}. With the shaft eliminated, there is also no detrimental stress caused by shrink-fit. This flywheel type requires specialized magnetic bearing and control that does not rely on a shaft\cite{Li2018}. A thin rim flywheel (shell flywheel) can also achieve a theoretical limit of 0.5. It is more compact since many system components like bearings and M/G can be integrated inside the rotor. A shell flywheel also requires specialized MB and MG\nomenclature{MG}{Motor generator} since it has no shaft \cite{Kailasan2014,Li2019}.

\subsubsection{Compact and integrated FESS}\label{compact}
Researchers have sought different means to reduce the system component and make FESSs more compact. Such design is vital to transportation applications, which are sensitive to space and weight. Kailasan et al. \cite{Kailasan2014} give the design for a 1 kWh inside-out integrated FESS. The rotor is shell-like and made of steel/composite materials. It allows the other components to be installed inside the shell flywheel. Many researchers also \cite{Filatov2012b,Ren2019,Hemenway2020} focus on the design of combined radial and axial magnetic bearings that bring the three-bearing configuration down to a two-bearing configuration. Recently, Li et al. \cite{Li2017h,Li2021ee} present a combination magnetic bearing for a shaftless flywheel. The single magnetic bearing can provide full levitation control \cite{Li2018}. Basaran et al. \cite{Basaran2019NovelControl} present radial repulsive magnetic bearings that reduce power consumption with less complexity. Andriollo et al. \cite{Andriollo2020DesignSystems} discuss an integrated, axial hybrid magnetic bearing with a steel flywheel. In \cite{Ershad2019}, an active electromagnetic slip coupling is developed to make a more compact and cost-effective flywheel-based powertrain. A bearingless electric machine, which is also reviewed in \ref{bearingless}, can act as the magnetic bearing and motor-generator at the same time, making the system more compact. For example, Yang et al. \cite{Yang2021} propose a bearingless flywheel motor to achieve a high integration level for vehicle applications. \par
Magnetic gears also make the FESS more compact by reducing the need for extra power electronics. Ricardo \cite{SpinningWheel2015} developed a FESS with magnetic gear for automotive application. As depicted in Fig.~\ref{typicalflywheel}, the system is compact and free of extra power electronics. In \cite{Andriollo2017}, an axial flux magnetic gear is designed to directly couple a FESS with a motor for recharging a heavy-duty electric bus. In general, more studies are needed to understand how the magnetic gear can meet the power, torque, speed, and efficiency requirements for FESSs. 

\subsubsection{New materials}
Apart from steel and carbon-fiber-based composite, some interesting proposals use new materials. One of the most promising materials is Graphene. It has a theoretical tensile strength of 130 GPa and a density of 2.267 g/cm3, which can give the specific energy of over 15 kWh/kg, better than gasoline(13 kWh/kg) and Li-air battery (11 kWh/kg),  and significantly higher than regular Li-ion batteries. In \cite{Kane2010LevitatedTrap}, graphene flakes are levitated and spun at rotational speeds of up to 60 million rpm. Unfortunately, it is unclear how the energy can be harvested. Sandia National Lab \cite{Timonthy-2016,Boyle2014} is working on improving flywheel energy density with Graphene to increase the flywheel's strength.  Circosta et al. \cite{Circosta2018} present a semi-hard magnetic FeCrCo 48/5 rotor that enables the use of bearingless hysteresis drives. Martin et al. \cite{Martin2016a} developed a new magnetic composite material that can be used for magnetic bearing and the rotor shaft. Magnetic permeability, saturation magnetism, mechanical stiffness, tensile elasticity, and electrical resistivity are considered.
The use of new materials, both in flywheel rotor and subsystems like the magnetic bearing, will enable the FESS to reach higher specific energy with a lower cost. Ideal materials for FESS rotor should have the following properties:
\begin{itemize}
  \item high tensile strength
  \item low cost
  \item low density
  \item magnetically permeable
  \item recyclable
  \item predictable fatigue life
  \item easy to monitor stress crack
\end{itemize}

Several less-proven concepts use low-cost materials. Energiestro \cite{Energiestro} promotes a flywheel made of concrete, claims that it "will decrease by a factor of ten the cost of energy storage." Similarly, Velkess \cite{velkess} has proposed a flywheel made of e-glass. However, both materials have very low tensile strength, it is not clear how they can be competitive in terms of costs and performance.

\subsubsection{Flywheel loss, failure modes, and containment}

Considering that Li-ion batteries have a low self-discharge rate, reducing the standby loss is crucial for making FESSs competitive \cite{Pullen2019TheStorage}. FESS losses come from the rotor (windage loss), the electric machine (core loss, copper loss), the AMB (eddy current loss and hysteresis loss), and the converter. There is some research activity on the standby and operational loss of FESS \cite{XJL2018-TAMU,Amiryar2020AnalysisSystems,Gengji2016RotorSupply}. But most of them focus on windage loss or motor-related losses. There are fewer works on AMB or converter-related loss. Also, there is a lack of work on loss-reducing methods \cite{Santra2019MultiphysicsStrategies}. 

More importantly, flywheels must be kept free from failures, which could end catastrophically \cite{blast-2016}. There is some work in failure mode analysis and prognosis. But most of them focus on the composite rotor \cite{Wang2018ProgressiveStorage,Chen2013ProgressiveTheory} or traditional mechanical bearings \cite{Cheng2019AAutomaton}. As flywheels are such complicated electro-magneto-mechanical devices, it is necessary to include other components and systematically investigate the failure mode and its containment. Recently, Buchroithner et al. \cite{app8122622} developed a test rig for the systematical investigation of burst containments under rotor failure. The goal is to investigate the performance of different containment structures and materials systematically.

\subsection{New Application}
\subsubsection{Energy-saving/harvesting}

\begin{figure}[bt]
\centering
\includegraphics[width=8cm]{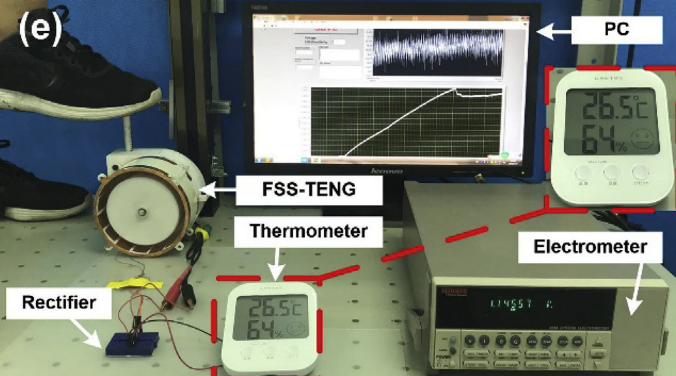}
\caption{Harvest foot motion energy to a flywheel, from \cite{Yang2019}}\label{flywheel-Tengs}
\end{figure}

Many of the industrial devices repeat certain motions. For example, a crane/truck lifts a heavy object and relocates it to a different place. A robot arm follows the planned motion trajectory and accelerates and decelerates to meet the speed and acceleration profiles. If the energy during these repeated motions can be harvested and reused for the next cycle, the efficiency can be improved. These charge/discharge cycles frequently occur with high power requirements, making the FESS a good candidate. We have noticed some commercial products deployed for large industry devices such as cranes \cite{Ahamad2019EnergySystems,Kermani2018}. However, there is less research work for smaller devices like robot arms \cite{Gale2015},  automation devices, and tooling machines. As for energy harvesting, Yang et al.\cite{Yang2019} present an interesting approach of storing energy harvest from triboelectric nanogenerators (Tengs) in a flywheel so that it can capture intermittent excitation (depicted in Fig.~\ref{flywheel-Tengs}). Such integration opens a new opportunity for FESSs. Future research topics in this area include the following:
\begin{itemize}
\item energy source and FESS integration topology
\item control scheme for optimal energy-saving and harvesting
\item cooperation strategy between multiple FESSs
\end{itemize}

\subsubsection{Energy buffer and hybrid storage system}
\begin{figure}[bt]
\centering
\includegraphics[width=8cm]{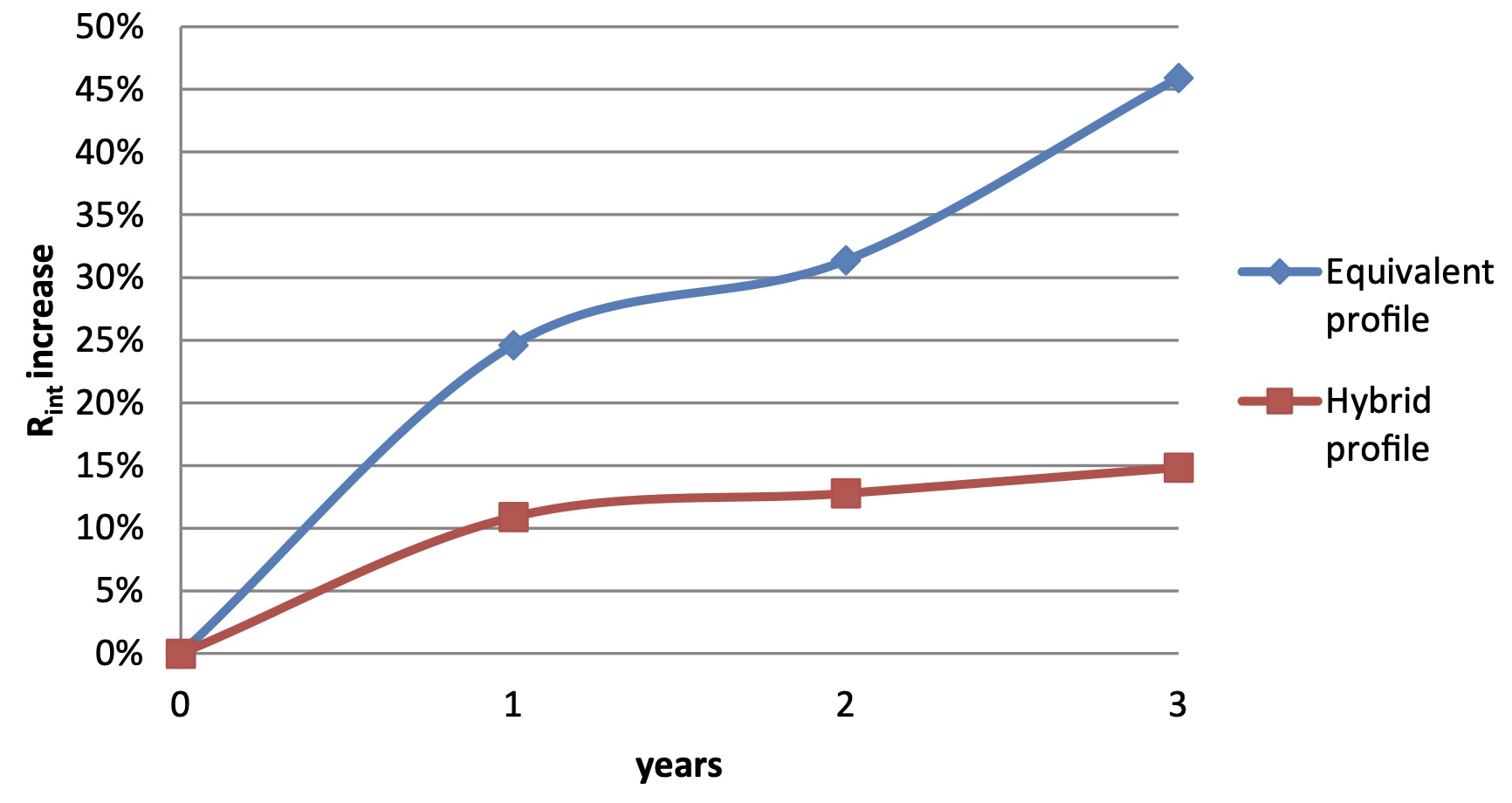}
\caption{Increase for LFP internal resistance in hybrid and non-hybrid configurations, from \cite{BARELLI2019937}}\label{life_gain}
\end{figure}

An excellent analogy for the relationship between flywheels and Li-ion batteries is the computer's memory architecture. A computer has multiple layers of memory devices. Fast memories such as cache and RAM (random access memory) are similar to FESS: fast-responsive and higher power/speed ratings. The slower device such as hard drives offers abundant storage at a low cost, similar to Li-ion batteries. Therefore it makes sense for an energy storage system to use a cascaded architecture that incorporates different technologies. The FESS should act as a buffer layer to provide a high-quality power output. In the meantime, it protects the batteries from being regularly charged/discharged so that the battery life is prolonged. This approach increases performance and lowers overall costs. For example, Barelli et al.\cite{BARELLI2019937} investigated the effects of integrating FESS with battery packs. As depicted in Fig.~\ref{life_gain}, the hybrid configuration greatly slowed the battery aging process by a factor of 300\%. 

While many papers compare different ESS technologies, only a few research \cite{Ramli2015EconomicStorage,BARELLI2019937} studies design and control flywheel-based hybrid energy storage systems. Recently, Zhang et al. \cite{ZHANG2018121} present a hybrid energy storage system based on compressed air energy storage and FESS. The system is designed to mitigate wind power fluctuations and augment wind power penetration. Similarly, due to the high power density and long life cycles, flywheel-based fast charging for electric vehicles \cite{Sun2016b,Ren2020,8433740} is gaining attention recently. Other advantages of flywheel-based supercharging include operability
under low/high temperatures, state-of-charge precision, and recyclability \cite{Xfkurlwkqhua}.

\subsubsection{Beyond energy storage}\label{att-control}
FESS has a unique advantage over other energy storage technologies: It can provide a second function while serving as an energy storage device. Earlier works use flywheels as satellite attitude-control devices. A review of flywheel attitude control and energy storage for aerospace is given in\cite{Lappas2009SurveySystems}. Superconducting magnetic bearings \cite{Tang2012} are proposed for satellite attitude control. In \cite{Mehedi_2017}, a full state-feedback control method is proposed to increase the satellite attitude performances. Simulation shows the attitude accuracy can be controlled up to 0.001\textdegree.

Furthermore, flywheel attitude control has been applied to robotics and vehicles. Hockman et al. \cite{Hockman2017DesignBodies} propose three spinning flywheels to internally-actuator a rover for low gravity planetary bodies. In \cite{Lee2018DetectionRobot}, a flywheel for balancing control of a single-wheel robot is presented. In \cite{Chu2017DesignPendulum}, two flywheels are used to generate control torque to stabilize the vehicle under the centrifugal force of turning. 

\section{Conclusion}
In this paper, state-of-the-art and future opportunities for flywheel energy storage systems are reviewed. The FESS technology is an interdisciplinary, complex subject that involves electrical, mechanical, magnetic subsystems. The different choices of subsystems and their impacts on the system performance are discussed. Owing to its unique advantages, many different FESS systems have been built and applied to a wide range of applications, including renewable energies, transportation, utilities, and more. This review focuses on the recently developed FESSs, such as the utility-scale and low-cost steel flywheels. Finally, we identify the future development for the FESS technology. The use of new materials and compact designs will increase the specific energy and energy density to make flywheels more competitive to batteries. Other opportunities are new applications in energy harvest, hybrid energy systems, and flywheel's secondary functionality apart from energy storage.

\bibliography{references}

\end{document}